# Enhanced high-temperature performance of GaN light-emitting diodes grown on silicon substrates

**Hyun Kum[1], Namsung Kim[1], Young Hwan Park[1], Joosung Kim[1], Daemyung Chun[1], Jongsun Maeng[1], Yuseung Kim[1], Jun-Youn Kim[1,\*], and Young Soo Park[1]**

[1] Advanced Development Team, LED Business
Samsung Electronics
Yongin-si, Gyeonggi-do
446-811, South Korea

**Dong-Pyo Han[2], Dong-Soo Shin[3], and Jong-In Shim[2]**

[2] Department of Electronics and Communication Engineering
Hanyang University, ERICA Campus
Ansan-si, Gyeonggi-do
426-791, South Korea

[3] Department of Applied Physics and Department of Bionanotechnology
Hanyang University, ERICA Campus
Ansan-si, Gyeonggi-do
426-791, South Korea

## Abstract

We compare the temperature dependence of optical and electrical characteristics of commercially available GaN light-emitting diodes (LEDs) grown on silicon and sapphire substrates. Contrary to conventional expectations, LEDs grown on silicon substrates, commonly referred to as GaN-on-Si LEDs, show less efficiency droop at higher temperatures even with more threading dislocations. Analysis of the junction temperature reveals that GaN-on-Si LEDs have a cooler junction despite sharing identical epitaxial structures and packaging compared to LEDs grown on sapphire substrates. We also observe a decrease in ideality factor with increase in ambient temperature for GaN-on-Si LEDs, indicating an increase in ideal diode current with temperature. Analysis of the strain and temperature coefficient measurements suggests that there is an increase in hole transport efficiency within the active region for GaN-on-Si LEDs compared to the LEDs grown on sapphire, which accounts for the less temperature-dependent efficiency droop.

Electronic mail: junyoun.kim@samsung.com.

Energy-efficient solid-state lighting based on light-emitting diodes (LEDs) has been actively developed over the past several decades, promising highly efficient lighting solutions in all areas that require illumination such as mobile devices, consumer electronics, vehicle headlights, and residential/industrial lighting. Its success has been internationally recognized, and is undoubtedly changing the landscape of lighting in the positive direction [1]. Gallium-nitride (GaN) has been the material of choice for white-light applications due to its high efficiency, robust reliability, and the fact that it can be grown on a wide range of substrates. The dominant substrate used today is sapphire, but active research is being done on other substrates such as silicon, silicon carbide (SiC), glass, and bulk GaN [2-11]. There are advantages and disadvantages of each substrate, but sapphire and silicon substrates are optimal for maximizing the cost-to-lumen ($/lm) ratio due to the wide availability of cheap, large-area substrates. One minor drawback of solid-state lighting is that the quantum efficiency tends to decrease as a function of current density (J-droop) and temperature (T-droop). The exact physical mechanism of efficiency droop as a function of current density in GaN LEDs has been a hotly debated topic over the past few years. There have been reports suggesting that Auger recombination may be the main culprit [12-14], but the exact mechanism is still very controversial, as other groups show that electron overflow may be the dominant factor [15, 16], or both depending on the current density [17]. An explanation that focuses on the saturation of the radiative recombination rate as an inherent cause and the subsequent increase in nonradiative recombination rates has also been proposed [18]. Efficiency droop as a function of temperature, however, has not been studied as thoroughly, especially in the high-temperature regime. The difficulty in analyzing temperature-dependent droop characteristics lies in the fact that apart from ambient temperature, the junction temperature must also be accounted for, which may vary depending on a variety of factors such as epitaxial structure, defect density, chip geometry, packaging, and operating current

density. Most LEDs generate heat during operation, and for mid- to high-power applications, the junction temperature may exceed the ambient temperature by up to hundreds of degrees at high current densities. As such, it is crucial to identify the factors that contribute to thermal droop since it is also closely related to the degradation, reliability, and lifetime of the device [19, 20]. In this study, we compare the thermal-droop characteristics of LEDs grown on sapphire and (111) silicon substrates to identify the mechanism affecting the radiative efficiency during high current and temperature operation.

For our study, two types of samples were prepared. Sample A and sample B denote LEDs grown on sapphire and silicon substrates, respectively. Both samples were grown via metal-organic chemical vapor deposition (MOCVD). An identical LED structure was grown for each sample, consisting of a Si-doped n-GaN layer for electron injection, an InGaN/GaN strain relaxation layer, five-pairs of multiple-quantum-wells (MQWs) active layer, a Mg-doped AlGaN electron blocking layer (EBL), and a Mg-doped p-GaN layer for hole injection. The first two quantum wells were grown thinner to prevent excessive degradation in the crystallinity of the quantum wells closer to the p-side. The majority of the radiative process is known to occur in the last quantum well. One minor difference between samples A and B is that sample B has a 150-nm-thick AlN nucleation layer and an AlGaN multi-layer buffer grown on top of the Si substrate to relieve stress caused by the lattice mismatch between GaN and Si [21,22], which is a common solution that makes silicon substrates viable for GaN growth. The threading dislocation densities (TDDs) of samples A and B were estimated to be ~$1 \times 10^8$ and ~$5 \times 10^8$ cm$^{-2}$, respectively. Due to different buffer structures between GaN and the substrate, sample B has natively more extended defects than sample A. After growth, both samples were made into identical vertical thin-film (VTF) LEDs with an area of $1 \times 1$ mm$^2$ using conventional fabrication and wafer-bonding techniques. The LEDs were then packaged and wire-bonded without encapsulation for electrical and optical characterization. Both samples had electroluminescence (EL) wavelengths of ~443-444 nm (blue) at a bias current of 350 mA.

The external quantum efficiency (EQE) and thermal droop characteristics of samples A and B were measured using an integration sphere as function of temperature. Measurement results only at 25 and 85 ˚C are shown for clarity. Consistent ambient temperature was maintained between, and during, measurements. The light extraction efficiency (LEE) can be assumed to be constant for both samples due to identical device geometry and packaging; LEE is also independent of temperature and current density [23,24]. Thus, the relative EQE trends of samples A and B can be interpreted as the internal quantum efficiency (IQE) trends of both samples, since EQE = IQE × LEE. Figure 1 (a) and (b) show the EQEs at 25 and 85 ˚C as a function of current. The efficiency at lower current densities is usually associated with recombination at nonradiative defect centers, i.e. Shockley-Read-Hall (SRH) recombination [25]. The SRH recombination typically saturates at higher current densities, but the maximum IQE is lowered as a consequence. The reduction of IQE at higher current densities is attributed to either Auger recombination or carrier overflow, as mentioned earlier. As expected, the efficiency of sample A is higher than that of sample B at 25 ˚C, but surprisingly, the EQE of sample B becomes higher at an elevated temperature of 85 ˚C.

The changes in forward voltage ($V_f$) and peak wavelength ($W_p$), both at 350 mA, were also measured, as shown in Fig. 2 (a) and (b). Sample A shows a larger redshift and smaller $V_f$ drop at higher temperatures compared to sample B. The redshift is a consequence of bandgap shrinkage while the $V_f$ drop is associated with decrease in resistance and ease of carrier diffusion due to high energy carriers. The $V_f$ drop can also be influenced by an increase in Mg activation and hole injection efficiency from the p-GaN layer [26]. Figure 2 (c) shows the T-droop (25–85 ˚C) as a function of current. The data was collected after thirty seconds of DC current biasing to allow the device to reach thermal equilibrium, where the radiative output and forward voltage values were independent of biasing duration. Surprisingly, sample B shows superior T-droop properties for all current values except at the lowest level (< 10 mA). This is an unexpected result since it has been reported that the higher TDDs lead to worse T-droop [27]. We will discuss the reason for these behaviors in detail later in this work. The $W_p$ shift as a function of current at 25 ˚C also shows a contrasting response for samples A and B (Fig. 2(d)). For sample A, blueshift initially occurs as current increases, followed by redshift. This redshift is not observed for sample B. The expected response is blueshift only, typically attributed to the screening of the piezoelectric field (PF) at low current densities, followed by state filling at high currents. The redshift of the peak wavelength indicates that there is relatively more heating at the junction, resulting in bandgap narrowing. Consequently, the difference in junction temperature is investigated for both samples.

Although the external ambient temperature can be controlled, the internal temperature may depend on intrinsic factors, such as the junction temperature ($T_j$) and thermal resistance of the packaging. Since identical packaging process was used, $T_j$ can be differentiated for each sample. The junction temperature at any operating condition can be deduced by measuring the temperature coefficient (TC), which is found by calculating the change in $V_f$ as a function of temperature ($\Delta V_f/\Delta T$), where the duration of the sensing current is short and the amplitude is small enough so that the junction may not be heated [28]. We used a sensing current of 10 mA and a pulse duration of 9 ms for the measurements. The measured results are shown in Fig. 3. A clear difference in the slope is observed, where the slope corresponds to the TC. Values of ~-0.92 mV/˚C and -1.5 mV/˚C are obtained for samples A and B, respectively. It is then possible to estimate the relative increase in junction temperature between 25 and 85 ˚C at an operating current of 350 mA by measuring the change in $V_f$ at these conditions ($V_{f\_25˚C}$, $V_{f\_85˚C}$). $V_f$ drops of 86.12 mV and 112.1 mV were measured for samples A and B, respectively. This corresponds to an increase in $T_j$ of approximately 94 °C for sample A and 75 °C for sample B, which means that sample A is intrinsically hotter than sample B by nearly 20 °C at an ambient temperature of 85 °C and a current bias of 350 mA. Consistent results were obtained for several samples. A difference in TC is typically attributed to the thermal conductivity of the active material. Since our device is composed of the same material and with the same epitaxial structure, the carrier dynamics within the active junction must be playing a role. We note that our VTF structure removes the native substrate, thus the difference in thermal conductivity between sapphire and silicon does not need to be considered.

The carrier transport mode in the junction can be predicted by measuring the temperature-dependent I-V characteristics (I-V-T). The I-V-T characteristics from 25 to 85 °C are shown in Fig. 4 (a) and (b). A contrast in current profile can be seen, where sample A has a definite change in slope in the low current regime (between 1.5 and 2.5 V), whereas sample B has a relatively small change in slope. This means that sample B has more tunneling carriers than sample A. The transport via tunneling is believed to have a very weak dependence on temperature. This also means that sample B has more energy states within the active region for the carriers to tunnel, which is reasonable considering the higher TDD of sample B. To investigate the effect of this difference on the diode behavior, we calculate the ideality factor for each sample. The ideality factor, $n_{ideal}$, is obtained by $n_{ideal} = (q/kT)(\partial \ln(I)/\partial V)^{-1}$, where q is the elementary charge, k is the Boltzmann's constant, I is the current, and V is the voltage, derived from the ideal diode equation [29]. An ideal diode would have an ideality factor of 1, which corresponds to the radiative recombination limited current. A value greater than 1 indicates non-radiative recombination limited current, such as SRH recombination and tunneling. The minimum values of $n_{ideal}$ for samples A and B change in different directions as a function of temperature as shown in Fig. 4 (c) and (d). For sample

A, as the temperature increases, the ideality factor also increases, which suggests that more carriers are lost due to SRH and nonradiative recombination processes. The minimum $n_{ideal}$ increases from 1.24 to 1.37 as temperature increases from 25 to 115 °C, respectively. In contrast, sample B shows an opposite trend: the ideality factor decreases, with the minimum $n_{ideal}$ decreasing from 1.46 to 1.32 as temperature increases in the same range, which implies an increase in radiative recombination processes.

To further clarify and elucidate this data, we calculate the radiative current for both samples. The radiative current, $I_{rad}$, is given by $I_{rad} = I \times IQE$, where I is the current obtained from the I-V measurements. As shown in Fig. 5 (a) and (b), the radiative current for sample A does not show much difference or even decreases at high currents with temperature, while for sample B it increases rather monotonically. Figure 5 (c) shows the radiative currents for both samples at 125 °C, clearly showing that more current is consumed radiatively for LEDs grown on silicon substrates.

We hypothesize that this could be due to two reasons. The first reason is the difference in activation of hole carriers. It is common knowledge that Mg activation in GaN grown by MOCVD is a challenge [30], where the active hole concentration is typically two orders of magnitude lower than that of the Mg dopant concentration. An increase in temperature activates the dormant Mg dopants, increasing the hole concentration and reducing resistivity. It has been reported that crystallographic strain could change the Mg activation energy [31], and in fact, the strain energy is much more compressive for GaN grown on sapphire than that for GaN grown on silicon. The strains were confirmed via X-ray diffraction (XRD) measurements for our samples (Fig. 6). A separate sample was prepared for hall measurements to observe the temperature dependence on the hole carrier concentration of the p-GaN layer grown on sapphire and silicon substrates. We do not observe any major difference in hole concentration as a function of temperature for both samples (Fig. 7), which allows us to conclude that the change in hole concentration is most likely not the reason for the difference in previously described temperature characteristics between samples A and B. An alternate reason must be found.

The second reason could be the dynamics of the carriers within the MQW active region. There is a difference in the PF caused by the lattice mismatch of the InGaN well and the GaN barrier. The PF is present in wurtzite III-nitride materials in the absence of external strain, and therefore an issue which reduces the electron and hole wavefunction overlap [32]. Fortunately, due to the relative tensile stress of GaN grown on silicon substrates compared to GaN grown on sapphire [33,34], the PF is reduced with respect to GaN grown on sapphire. This reduction not only increases the electron-hole wavefunction overlap, it also reduces the relative height of the GaN barrier for holes. This could be verified by measuring the current-dependent EL spectra at room temperature and high temperature. The results are shown in Fig. 8 (a)-(d). At 25˚C, the EL spectra for both samples show broadening of the full-width at half maximum (FWHM), consistent with band filling. We note that the EL measurements were done with pulsed signals, so effects of heating due to current flow can be neglected. At 125˚C, however, a peculiar behavior in the EL spectra can be seen for sample B. For sample A, there is no significant difference in the spectrum, whereas for sample B, a second peak with a shorter wavelength can be clearly seen emerging near 434 nm. We believe this peak is the result of holes being injected into the first two quantum wells with a thinner well thickness due to the lower barrier height, a result which we attribute to the relative tensile stress on the active region mentioned earlier. This roughly matches the difference in wavelength between the first two wells and the rest of the wells measured by photoluminescence. A double peak Gaussian fitting could be done for sample B, whereas it is impossible to do the same for sample A (see Fig. 9). These results allow a comprehensive modeling of the T-droop characteristics.

We go back to Fig. 2 (c), which embodies the essential physical mechanisms revealed in this study. The curve can be explained by three competing phenomena: (1) the activation of defect states within the forbidden gap, (2) carrier overflow, and (3) hole transport efficiency within the active region. At low currents (less than 10 mA), the dominant phenomenon is the activation of defect states within the forbidden gap. Thus, as the temperature is increased at low currents, the samples with higher defects show a larger power drop. Indeed, sample B which has a higher density of defects has a larger T-droop (~7.9%) at the low current region compared to sample A (~7.0%). In this current region, overflow is almost nonexistent and the hole transport efficiency is negligible. As the current is increased, electron overflow and hole transport start to dominate, whereas the defect states become saturated [35]. The T-droop for both samples improves in a moderate current region since the hole transport has a bigger effect on the output power than the overflow in this current range. However, as observed in the temperature- and current-dependent EL spectra in Fig. 8, sample B has a much superior hole transport efficiency across the active region, leading to a larger improvement in T-droop. This is also indirectly observed in the ideality factor analysis in Figs. 4 (c) and (d). At the highest current regime, both samples succumb to carrier overflow and T-droop deteriorates rapidly. Figure 2 (a) and (b) also support this model. Due to the higher defect density and superior hole transport, the GaN-on-Si LED shows a larger $V_f$ drop at higher temperatures for all current ranges. The improved hole transport also reduces electron overflow, leading to a cooler device, i.e., smaller TC, resulting in a smaller redshift due to the bandgap shrinkage, as evidenced in Fig. 1 (d). Also, although we do not have conclusive evidence, we hypothesize that a moderate amount of defects present in sample B prevent excessive carrier pileup and electron-electron scattering over the EBL layer, leading to a lower junction temperature as evidenced in Fig. 3. Unfortunately, a detailed discussion on this matter is beyond the scope of this work.

In conclusion, comparison of temperature-dependent efficiencies of GaN LEDs grown on silicon and sapphire substrates reveals that due to a relative tensile stress of the active layers for LEDs grown on silicon substrates, a reduction in PF lowers the barrier height for efficient hole transport, leading to superior T-droop characteristics in typical operating conditions. These results also signify that performance-wise, LEDs grown on silicon substrates may have a fundamental advantage of lower PF, and can be a competitive platform for mid- to high-power lighting applications.

**Figure captions**

**Figure 1** The EQEs of GaN LEDs grown on sapphire (sample A) and silicon (sample B) at (a) room temperature and (b) 85 ˚C.

**Figure 2** The changes in (a) $V_f$ and (b) $W_p$ at a current of 350 mA going from 25 to 85 ˚C as a function of current. (c) T-droop characteristics (25 - 85 ˚C) for samples A and B with respect to current. Regions I, II, and III indicate low, moderate, and high current regime, respectively. (d) $W_p$ shift at 25 ˚C for samples A and B with respect to current.

**Figure 3** $V_f$ as a function of temperature at low current bias (10 mA) and short pulse duration (9 ms) for sample A and B. The slope represents the TC of the LED.

**Figure 4** I-V characteristics for various temperatures for (a) sample A and (b) sample B. Calculated $n_{ideal}$ for various temperatures for (c) sample A and (d) sample B.

**Figure 5** The radiative current as a function of current for various temperatures for (a) sample A and (b) sample B. (c) A superimposed radiative current graph of sample A and B at 125 ˚C.

**Figure 6** Lattice constants of GaN grown on sapphire and silicon substrates with respect to free-standing GaN measured by XRD.

**Figure 7** Hole concentrations as a function of temperature measured by the Hall technique for samples A and B.

**Figure 8** Current-dependent EL spectra at 25 and 125 ˚C for sample A ((a), (b)) and sample B ((c), (d)), respectively.

**Figure 9** Gaussian fitting of the EL spectra shown in Fig. 8 for a current of 800 mA at 125 ˚C for (a) sample A and (b) sample B.

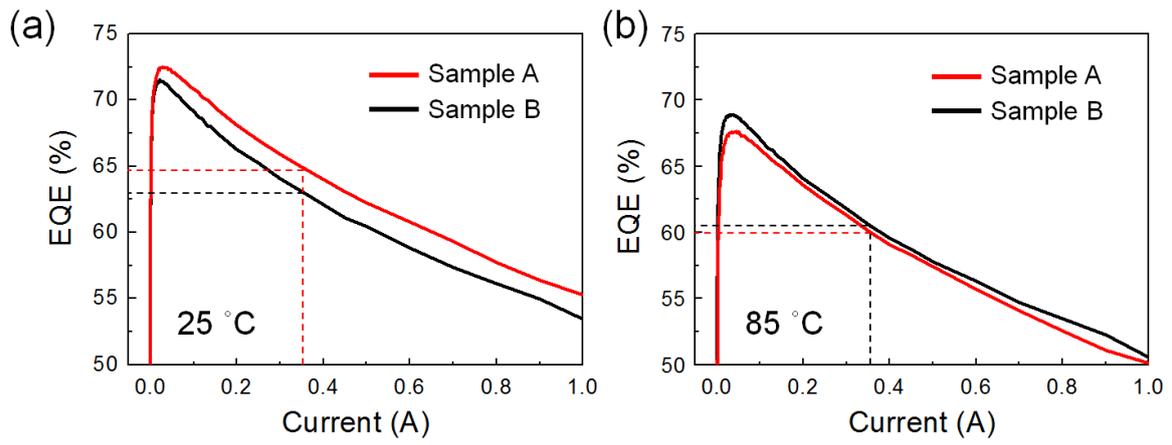

Figure 1 of 9

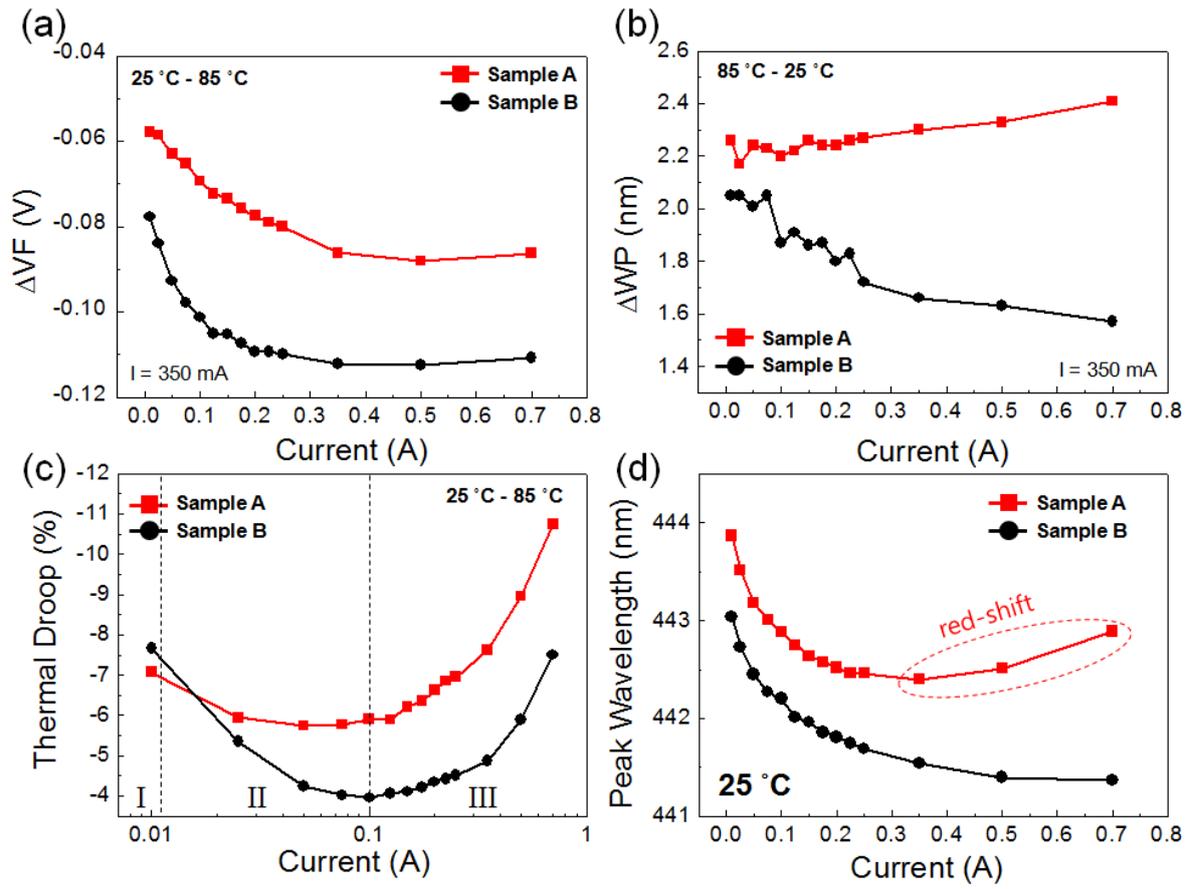

Figure 2 of 9

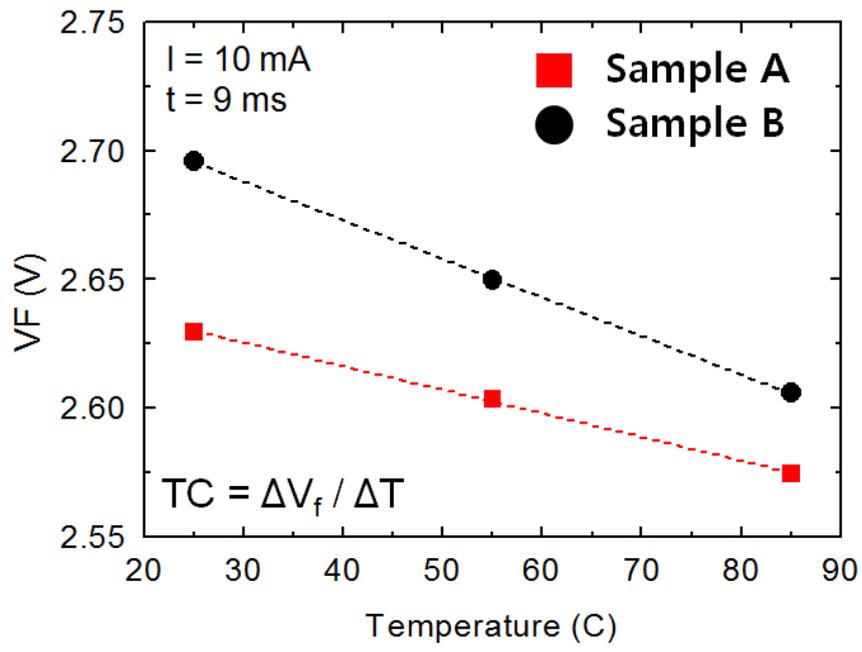

Figure 3 of 9

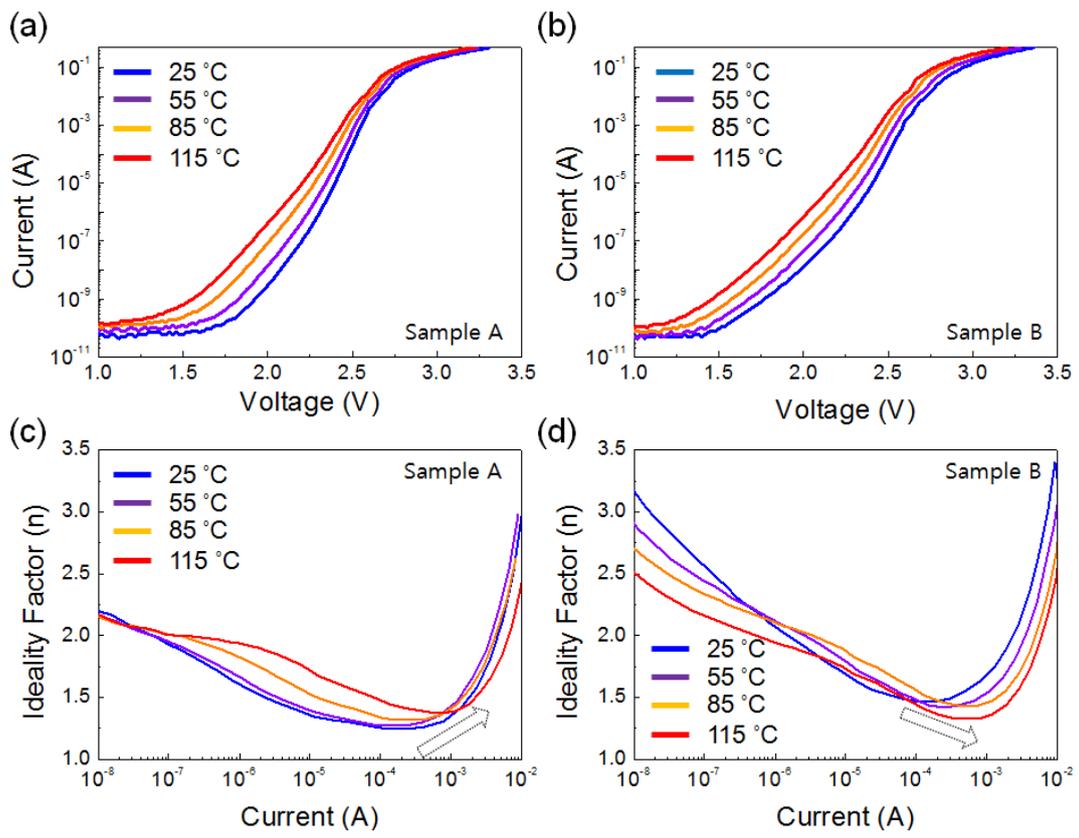

Figure 4 of 9

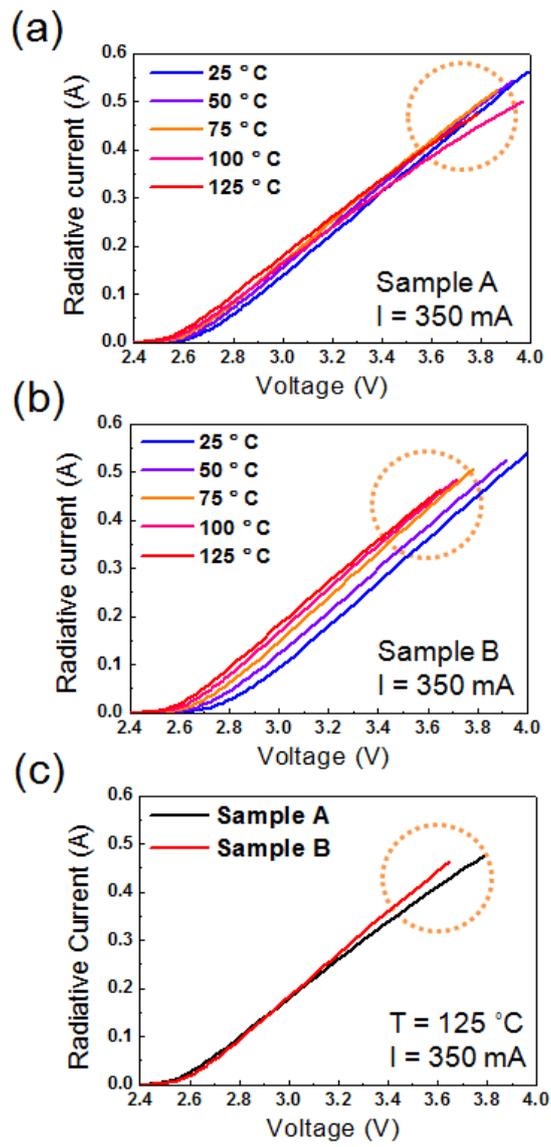

Figure 5 of 9

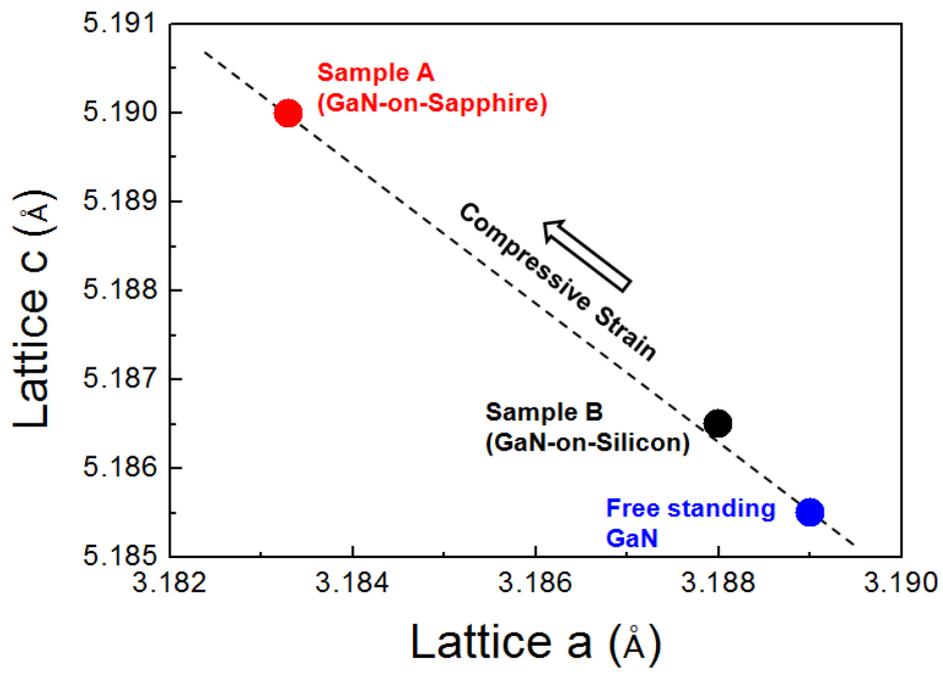

Figure 6 of 9

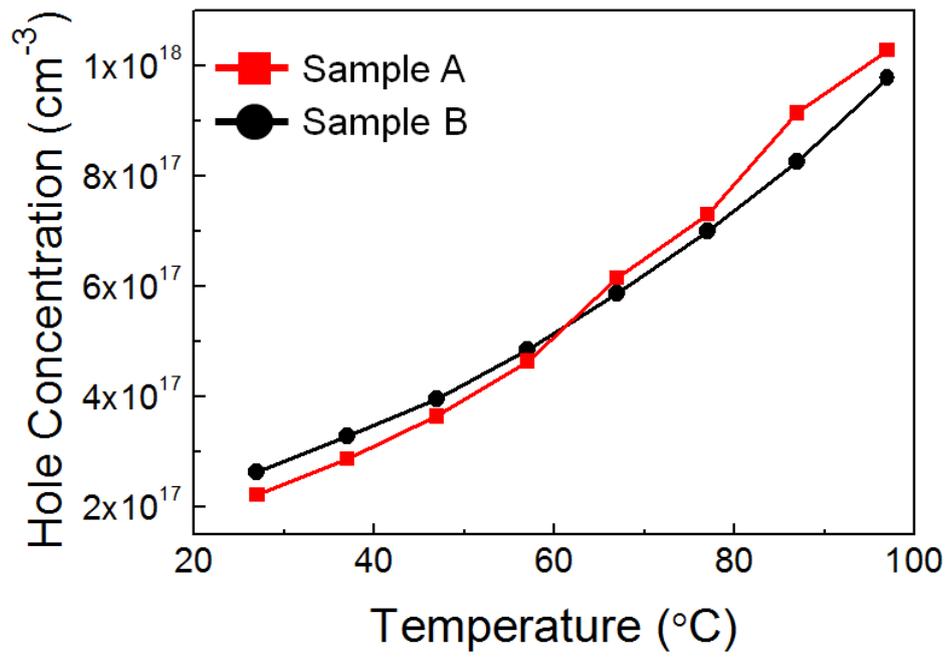

Figure 7 of 9

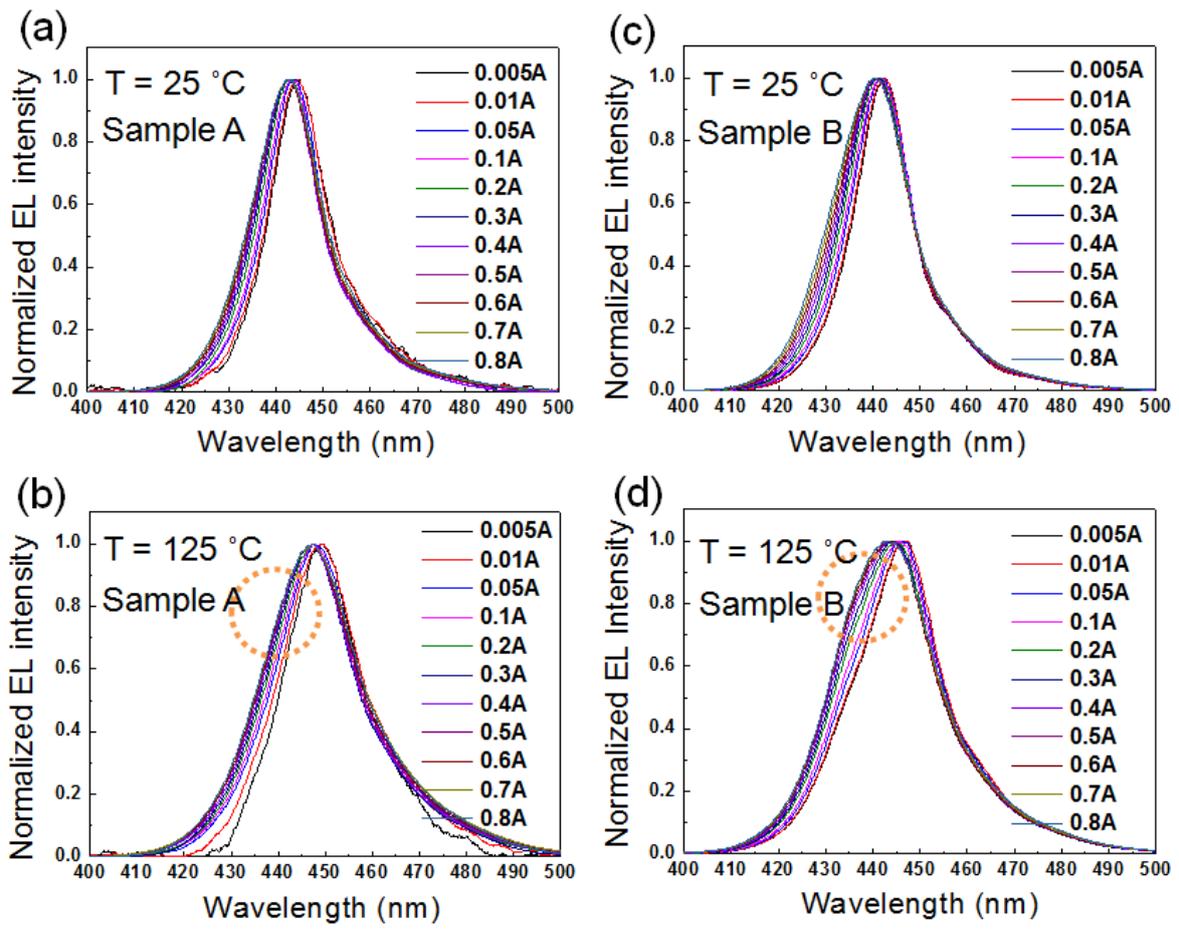

Figure 8 of 9

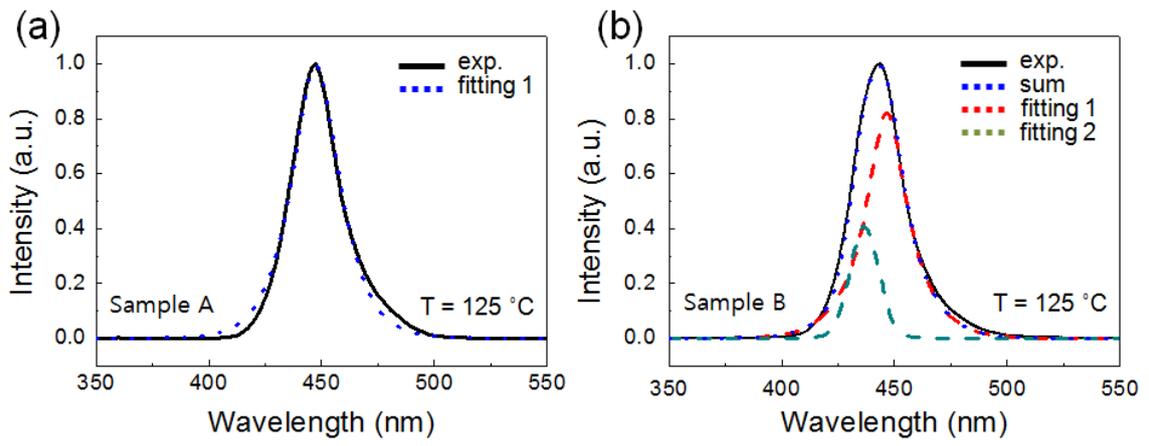

Figure 9 of 9